\documentclass[fleqn,12pt,twoside]{article}
\usepackage{espcrc1}
\usepackage{floatflt}
\usepackage{wrapfig}

\usepackage{graphicx}
\usepackage[figuresright]{rotating}

\newcommand{\AmS}{{\protect\the\textfont2
  A\kern-.1667em\lower.5ex\hbox{M}\kern-.125emS}}

\def\pt      {p$_t$}
\def\y       {y}
\def\cm      {~cm}
\def\gev     {~GeV}
\def\gevc    {~GeV/c}

\def\mev     {~MeV}

\def\piz     {$\pi^0$}
\def\gzepz   {${\gamma}Z{\rightarrow}e^-e^+Z$}
\def\pizgg   {$\pi^0\rightarrow\gamma\gamma$}
\def\etagg   {$\eta^0\rightarrow\gamma\gamma$}
\def\ele     {$e^-$}
\def\pos     {$e^+$}
\def\sqnn    {$\sqrt{s_{_\mathrm{NN}}}$}
\def\au      {$^{^{197}}$Au}
\def\auau    {\au+\au}
\def\pb      {$^{^{208}}$Pb}
\def\pbpb    {\pb+\pb}
\def\armGas  {10\% CH$_4$ and 90\% Ar}

\title{Measurements of Photon and {\piz} Production in Heavy Ion Collisions at RHIC}

\author{I.J. Johnson\address{Lawrence Berkeley National Laboratory, \\
One Cyclotron Road, MS 70R0319, Berkeley, CA 94720, USA} for the STAR collaboration\thanks{For the full author list and acknowledgements, see Appendix
"Collaborations" of this volume.}}

\begin{document}
\maketitle

\begin{abstract}
Inclusive transverse momentum spectra of photons and {\piz}s about mid--rapidity are presented for {\auau} collisions at {\sqnn} = 130\gev. Photon pair conversions have been reconstructed from charged tracks measured with the central Time Projection Chamber (TPC) of STAR. Contributions of the {\pizgg} decay in the inclusive photon spectra have been studied. It was found that this contribution begins to decrease at a transverse momentum of 1.65{\gevc} in the top 11\% most central centrality class.
\end{abstract}

\section{Introduction}
For more than 20 years, photons have been considered one of the most valuable probes of the dynamics and properties of the matter formed in heavy ion collisions \cite{shuryak1,thesis,WA98_direct_photons,directPhotonRecentReview}. In contrast to hadronic particles that have large interaction cross sections in dense matter, photons only interact electromagnetically and therefore have a longer mean free path. This path length is typically much larger than the transverse size of the hot dense matter created in nuclear collisions \cite{hadronGas}. Therefore, with high probability, photons will escape from the system undisturbed and preserve the history of their birth. Photons that radiate from and escape the systems created in heavy ion collisions provide direct information about the physical conditions under which they were created. Photons are produced in all stages of heavy ion collisions \cite{hadronGas,brem_space_time_evolution,flash,a1_meson,two_loop,two_loop2}, starting at the instance when the quarks and gluons of the opposing nuclei interact through to long lived electromagnetic decays. However, separating photons created in the earliest stages from other later and more copious production mechanisms is a current challenge.

In heavy ion collisions, electromagnetic decays of mesons are the dominant source of photon production. Alone the {\pizgg} and {\etagg} decays compose 97\% \cite{WA98_direct_photons} of the inclusive photon spectrum in {\pbpb} collisions at {\sqnn} = 17\gev. Thus, it is essential to know the contributions from these decays in order to understand the composition of the inclusive photon spectrum. The year 2000 design of the STAR experiment at RHIC has made both measurements of photon and {\piz} spectra possible. These spectra were then used to calculate contribution of the {\pizgg} decay in the inclusive photon spectrum as a function of \pt.

\section{Reconstructing Photons via Pair Creation \gzepz}
The dominant interaction process for photons traversing matter with energy above 10{\mev} is pair conversion, \gzepz. One technique of measuring photons is to reconstruct the pair conversions that occur in detector material. In STAR; parts of the Silicon Vertex Tracker, and the inner field cage and gas (\armGas) of the TPC were utilized as photon converters. This material lies within the main TPC, so that the charged particle daughters ({\ele} and {\pos}) that stem from conversions in this material will traverse the TPC's tracking volume. A three step process that
\begin{wrapfigure}[27]{r}{.38\textwidth}
\vspace{-10mm}
\resizebox*{.38\textwidth}{!}{\includegraphics*{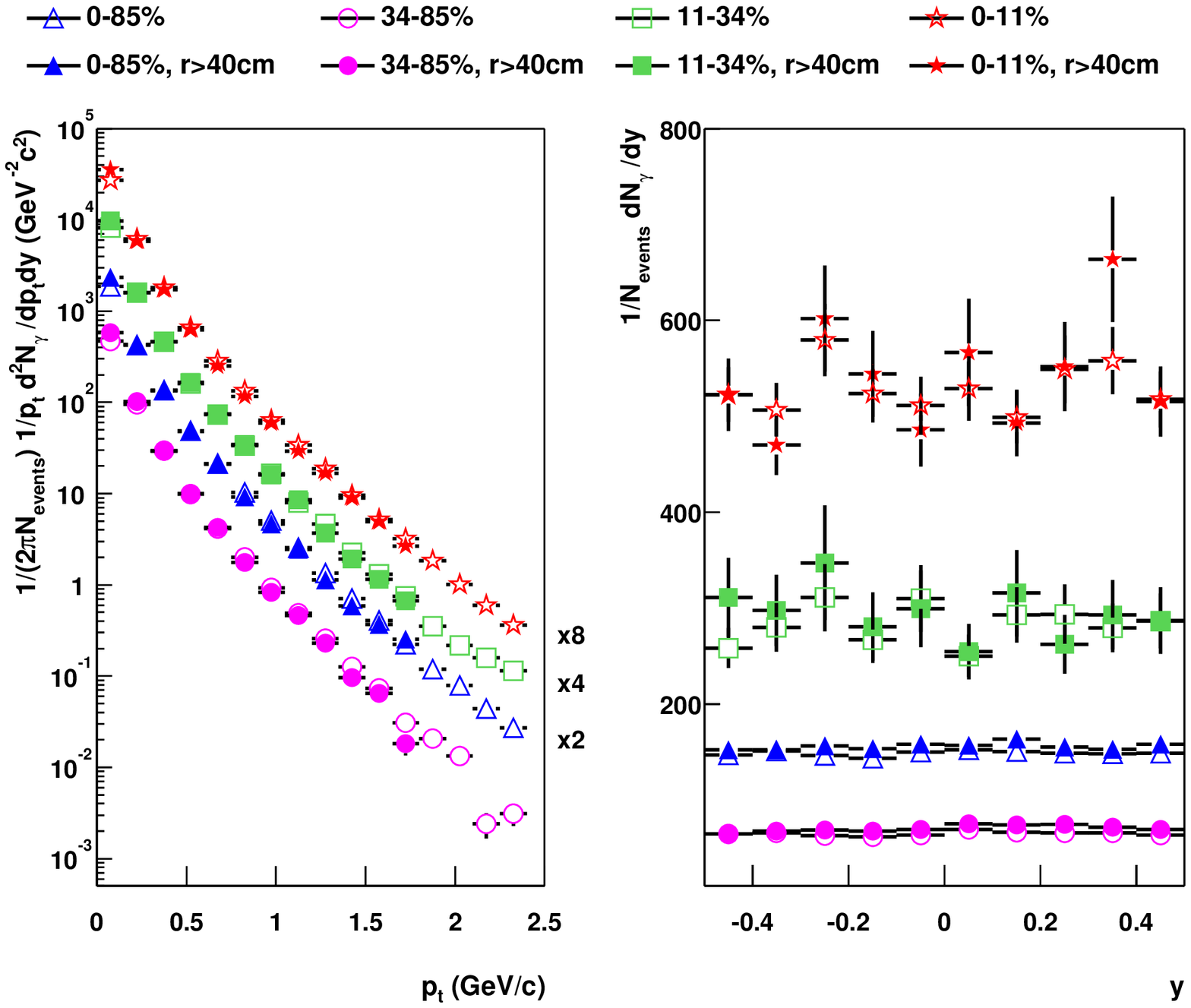}}
\vspace{-10mm}
\caption{Photon spectra about mid--rapidity, $|$\y$|<0.5$, for {\auau} collisions at {\sqnn} = 130{\gev}, $r_{xy}$$>$10{\cm} open symbols and $r_{xy}$$>$40{\cm} closed symbols. The errors shown are only statistical. An additional 7\% point--to--point systematic uncertainty has been estimated.}
\vspace*{-130mm}\hspace*{34mm}
\begin{minipage}[h]{40mm}
\begin{minipage}[t]{25mm}
\hspace*{-14mm}
\resizebox*{25.4mm}{!}{\includegraphics*{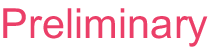}}
\end{minipage}
\begin{minipage}[b]{25mm}
\resizebox*{!}{4.6mm}{\includegraphics*{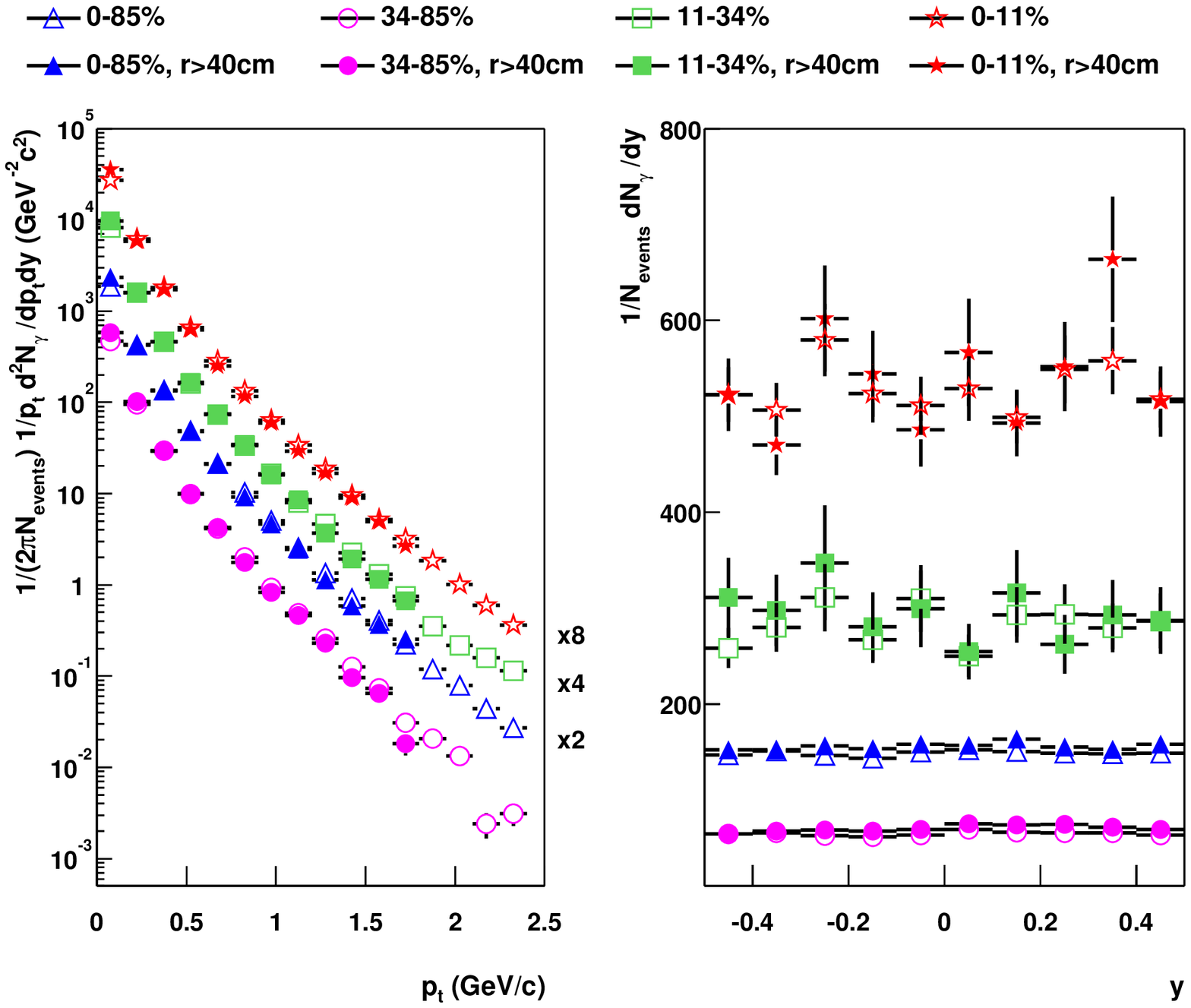}}
\resizebox*{!}{4.6mm}{\includegraphics*{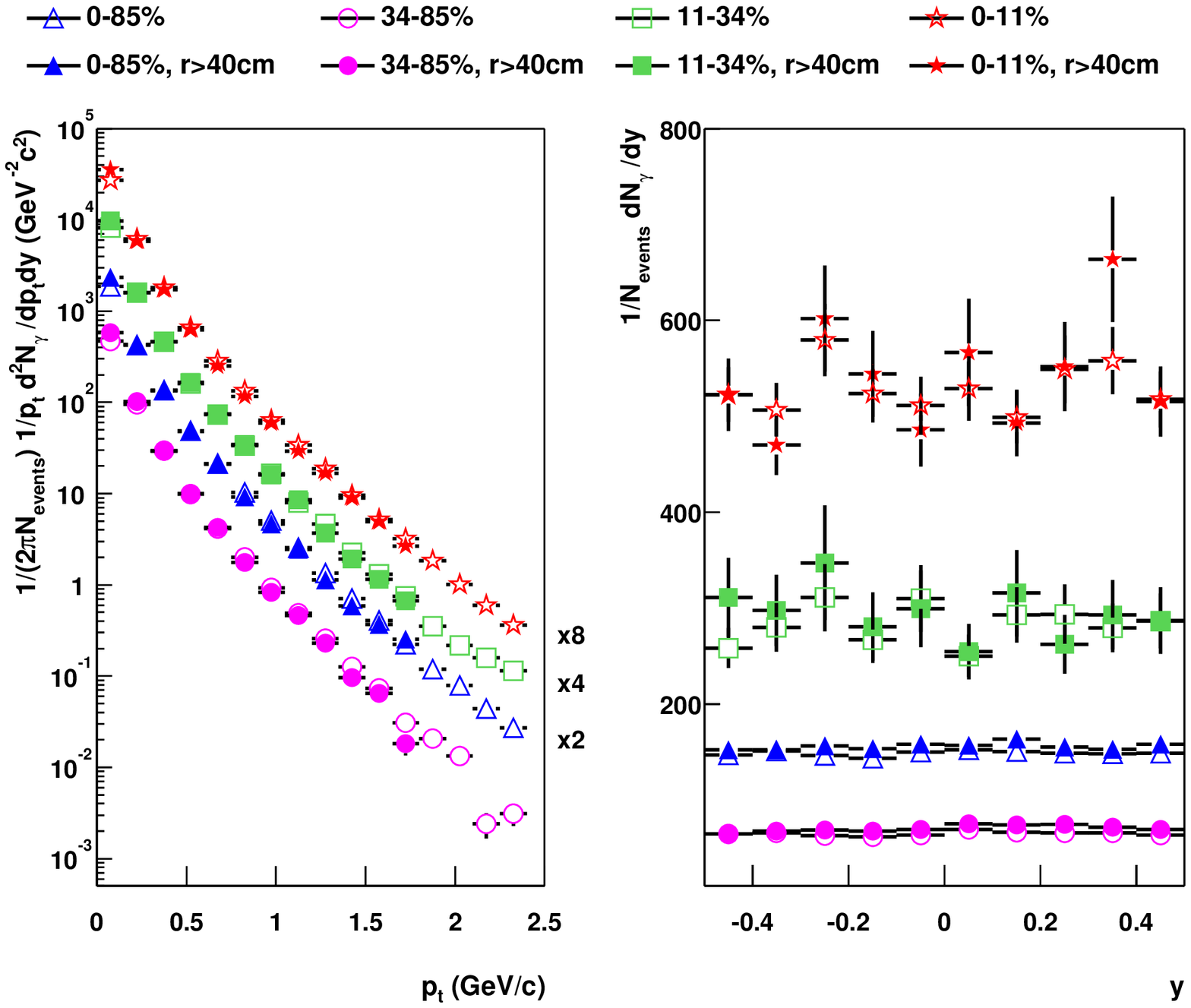}}
\resizebox*{!}{4.6mm}{\includegraphics*{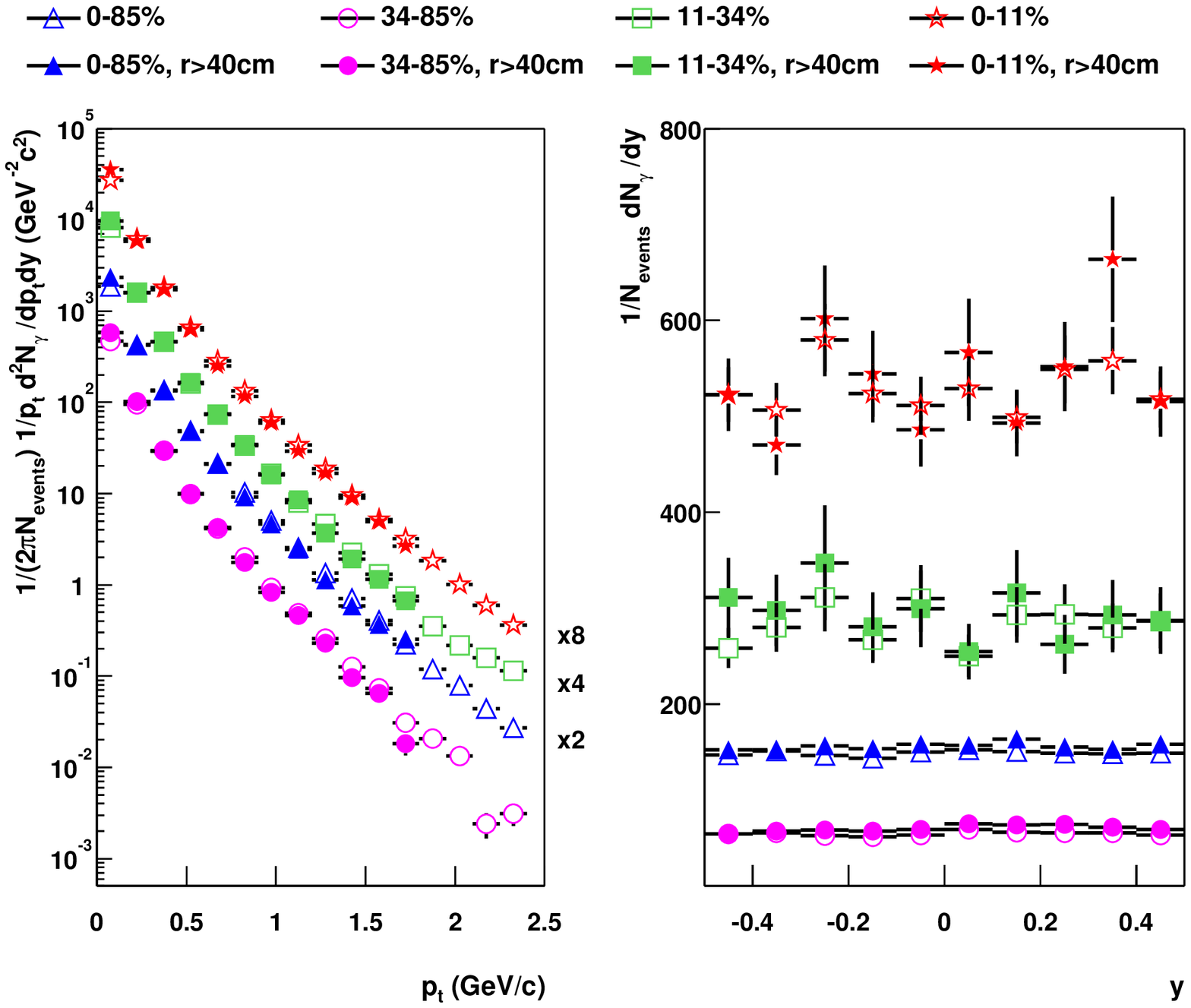}}
\resizebox*{!}{4.6mm}{\includegraphics*{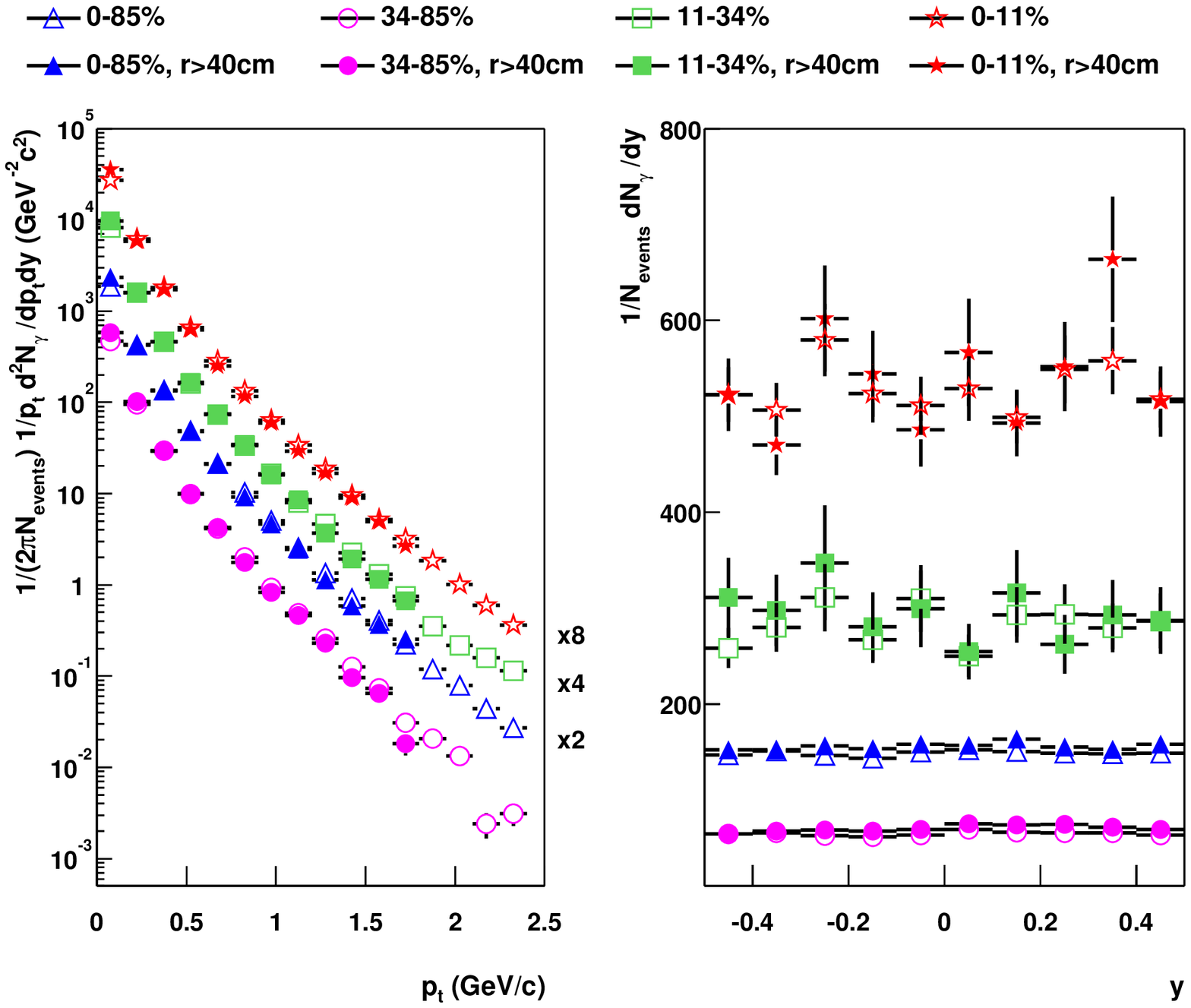}}
\end{minipage}
\end{minipage}
\label{fig:photonSpectra}\end{wrapfigure}
consisted of track, pair and primary photon selection cuts was used to reconstruct photon conversions from pairs of charged tracks measured by the TPC. Track cuts select potential daughters of photon conversions based on track geometry and loose dE/dx particle identification that only eliminated 2.275\% of the true electron daughters. Pairs of tracks were required to exhibit the unique topological signature of a photon conversion -- two tracks of opposite charge emerging from a secondary vertex with a small opening angle (${\approx}2m_ec^2/E_\gamma$ radians). Primary photon candidates were extracted by requiring the momentum vector obtained from the pair topology to point away from the primary vertex within an angular resolution of a few degrees.
The predicted energy loss as a function of momentum for electrons and positrons, dE/dx resolution of the TPC, and measured dE/dx and rigidity of daughters from primary photon candidates were combined into one variable, the dE/dx deviant. This variable is very powerful since it is independent of the daughter's momentum and ultimately independent of the parent photon's {\pt}. Raw yields of photons were extracted in slices of photon {\pt} through fits to distributions of the dE/dx deviant of the positive daughter. These distributions were fit with a Gaussian function plus an exponential function that described the background. Parameters of the exponential function were fixed and obtained through fits to distributions of the dE/dx deviant of positive tracks that came from pairs that lay in the outskirts of the geometric cuts, i.e. primarily background candidates. The daughter of negative charge was not chosen because, unlike positrons, electrons are present in detector material and processes like those that produce $\delta$--electrons pose as backgrounds. Efficiency and acceptance corrections were calculated with Geant 3.21 and applied to the raw yields. The Geant detector simulations included a realistic detector material map, though scatter and contour plots of the locations of photon conversions of real and simulated data revealed differences between these geometries. Most discrepancies were lessened through improvements to the simulation; but some areas, like those of wiring, could only be improved with more detailed approximations. Comparisons between real and simulated data confirmed that the inner field cage and gas of the TPC were of the right composition and density. The consistent conversion probability in this region ($r_{xy}$$>$40\cm) was used to calculate a scale factor which compensated for remaining uncertainties in the complete geometry ($r_{xy}$$>$10\cm). This scale factor was applied to the final all material photon spectra ($r_{xy}$$>$10\cm). Both the $r_{xy}$$>$40{\cm} spectra and the scaled $r_{xy}$$>$10{\cm} spectra are shown in Figure \ref{fig:photonSpectra}.

\section{Detecting {\piz}s}
\begin{wrapfigure}[20]{r}{100mm}
\vspace{-10mm}
\mbox{
\begin{minipage}[b]{100mm}
\begin{minipage}[b]{40mm}
\begin{minipage}[t]{40mm}
\resizebox*{\textwidth}{!}{\includegraphics*{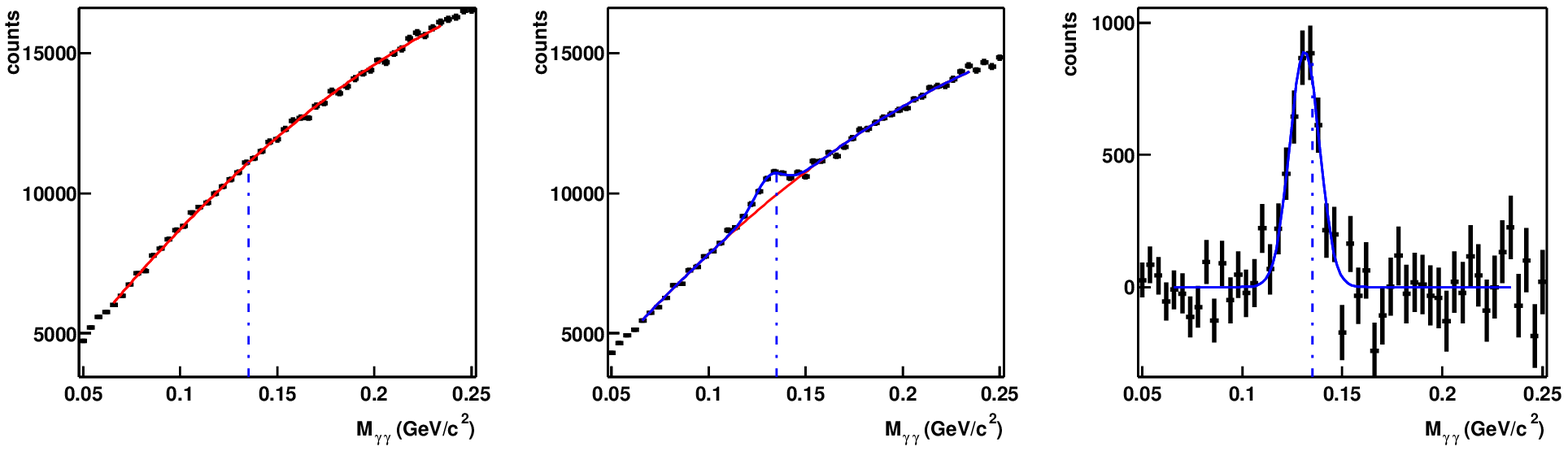}}
\end{minipage}
\begin{minipage}[b]{40mm}
\vspace{-2mm}
\resizebox*{\textwidth}{!}{\includegraphics*{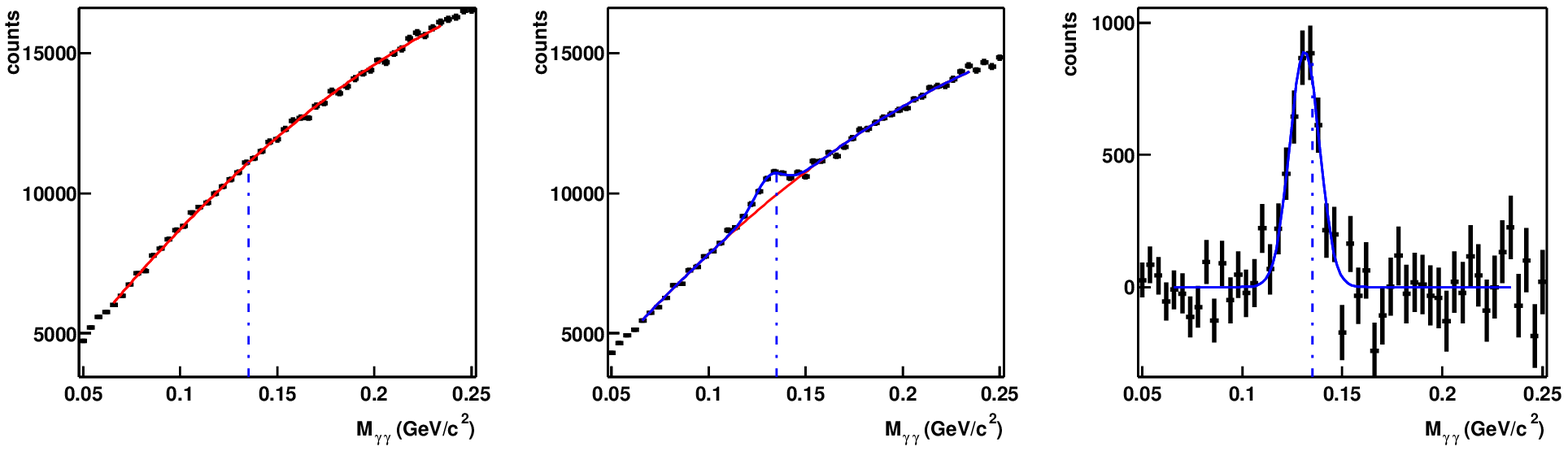}}
\end{minipage}
\end{minipage}
\vspace{-1.5mm}
\begin{minipage}[b]{60mm}
\resizebox*{\textwidth}{!}{\includegraphics*{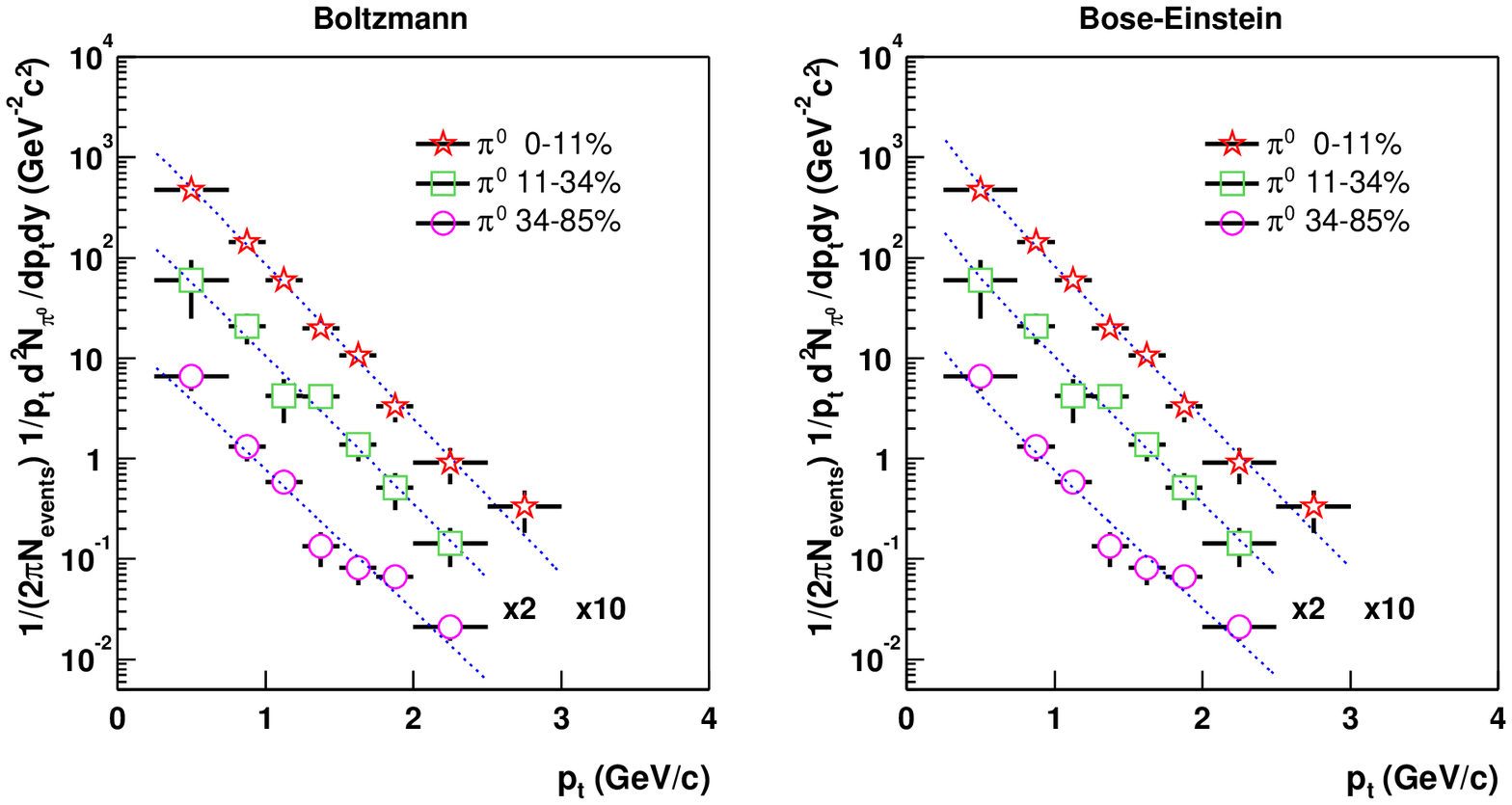}}
\end{minipage}
\end{minipage}
\vspace{-10mm}
}
\caption{Left: Two photon invariant mass distributions of photon pairs in {\auau} collisions at {\sqnn} = 130\gev, the top is before and the bottom is after background subtraction. Right: {\piz} spectra about mid--rapidity, $|$\y$|$$<$1, and dashed lines represent Bose--Einstein fits to the data points.}
\vspace*{-98mm}\hspace*{60mm}
\resizebox*{25.4mm}{!}{\includegraphics*{figures/preliminary.eps}}
\label{fig:pi0Spectra}
\end{wrapfigure}

Transverse momenta spectra of {\piz}s were measured through the {\pizgg} decay channel. Raw {\piz} yields were extracted by fitting the enhancement at the {\piz} mass in two photon invariant mass distributions, as shown in Figure \ref{fig:pi0Spectra}. These distributions were fit with a Gaussian function plus a second order polynomial to describe the background. The parameters of the second order polynomial were fixed and obtained by fitting invariant mass distributions of photon pairs that had one photon rotated by $\pi$ in azimuth. This procedure took
advantage of the $\pi$--azimuthal symmetry of the STAR TPC to preserve event characteristics like anisotropic flow, multiplicity and the location of the primary vertex. At the same time the rotation smeared and shifted the invariant mass of correlated photons. Raw yields as a function of {\pt} were obtained by slicing the invariant mass distribution into {\pt} bins. Efficiency and acceptance corrections were calculated with the realistic Geant simulations and applied to the raw spectra. Corrected spectra are shown in Figure \ref{fig:pi0Spectra}. Along with the shown point--to--point statistical errors, all spectra have a common uncertainty in the normalization of {$\pm$}40\%. The central data (0--11\%) has an additional uncertainty in the normalization of {$\pm$}19\%, due to the differing distributions of event vertex positions, resulting in a total uncertainty of {$\pm$}49\%.

\section{{\piz} Contributions in the Photon Spectra}
With both the inclusive photon and {\piz} spectra measured, the fractional contribution from {\pizgg} in the inclusive photon spectrum can be approximated. The {\piz} {\pt} spectra were fit to a Bose--Einstein function. These functional forms were used as the input distributions for a {\pizgg} decay simulator that generated resulting single photon spectra.
\begin{wrapfigure}[15]{r}{.4\textwidth}
\resizebox*{.4\textwidth}{!}{\includegraphics*{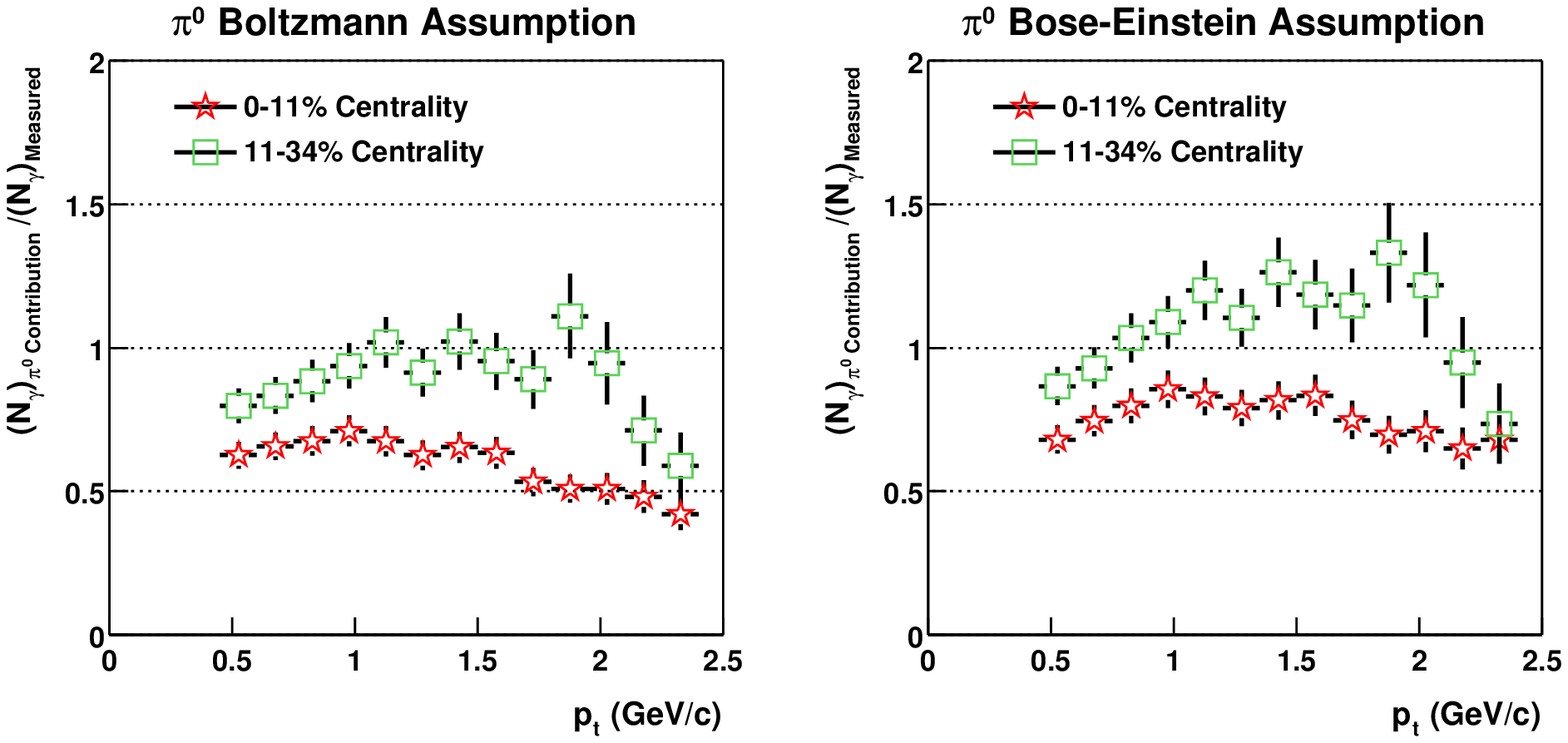}}
\vspace{-10mm}
\caption{Fractional {\pizgg} decay contributions in the inclusive photon spectra.}
\vspace*{-32mm}\hspace*{20mm}
\resizebox*{25.4mm}{!}{\includegraphics*{figures/preliminary.eps}}
\label{fig:photComp}\end{wrapfigure}
These generated spectra could be directly compared to the measured inclusive photon spectra, since both include efficiency and acceptance corrections. The fractional contributions under the Bose--Einstein assumptions for top 11\% most central and the 11--34\% centrality classes are shown in Figure \ref{fig:photComp}. This figure includes both the statistical and systematic uncertainties in the photon spectra, but the uncertainties in the normalizations of the {\piz} spectra have not been included. This leads to uncertainies in the relative and overall scales of the ratios, but does not affect the shapes. Above \pt=1.65{\gevc}, the contribution of the {\pizgg} in the inclusive photon spectra decreases in the 0--11\% centrality class. This change indicates an increase in contributions from other photon sources, possibly other electromagnetic decays or direct photons.

\section{Future Directions}
Accounting for the contribution from the {\pizgg} decay is the first step towards extracting direct photon yields. Contributions from other decays must be accounted for in order to understand the observed decrease that starts at 1.65{\gevc}. The production rates of other particles that decay via photon channels, like the \etagg, can be estimated with statistical models or directly measured experimentally. The energy resolution of photons detected via the pair conversion technique (${\Delta}E/E{\approx}$2\% at 0.5\gevc) makes the latter more feasible in the central {\auau} events at RHIC energies, which consequently have high photon multiplicities that result in large combinatoric backgrounds in two photon invariant mass distributions. A statically significant {\etagg} peak was observed in the 2000 data set at {\sqnn} = 130\gev. STAR's capability to measure photons through pair conversions has been illustrated through the results presented in this article. These results were extracted from the year 2000 data set that contains a factor of ten less statistics than the more recent 2001 data set. With the established tools to detect photons and the new data set, STAR in principle is able to measure photon, {\piz} and $\eta$ production. This will take measurements at RHIC one step closer to extracting the rates of direct photon production. Further details about the technique and the analysis can be found in \cite{thesis}.


\end{document}